\documentclass[twocolumn,amsmath,aps]{revtex4}
\usepackage{graphicx,color}
\usepackage{CJK}
\usepackage{bm}
\usepackage[hypertex]{hyperref}
\usepackage{float}

\newcommand{\be}{\begin{equation}}
\newcommand{\ee}{\end{equation}}
\newcommand{\bea}{\begin{eqnarray}}
\newcommand{\eea}{\end{eqnarray}}
\newcommand{\bsube}{\begin{subequations}}
\newcommand{\esube}{\end{subequations}}

\newcommand{\Eq}[1]{Eq.\,(\ref{#1})}
\newcommand{\Eqs}[1]{Eqs.\,(\ref{#1})}

\newcommand{\dg}{\dagger}
\newcommand{\la}{\langle}
\newcommand{\ra}{\rangle}

\newcommand{\nl}{\nonumber \\}

\newcommand{\gam}{\gamma}

\newcommand{\eps}{\epsilon}

\newcommand{\lam}{\lambda}

\newcommand{\beq}{\begin{equation}}
\newcommand{\eeq}{\end{equation}}
\newcommand{\beqn}{\begin{eqnarray}}
\newcommand{\eeqn}{\end{eqnarray}}
\newcommand{\bsub}{\begin{subequations}}
\newcommand{\esub}{\end{subequations}}

\newcommand{\ket}[1]{{\left| #1 \right\rangle }}

\newcommand{\ddg}{d^\dagger}
\newcommand{\fdg}{f^\dagger}

\begin{document}

\begin{CJK*}{GBK}{Song}

\title{Revisit the non-locality of Majorana zero modes and teleportation:
Bogoliubov-de Gennes equation based treatment}

\author{Xin-Qi Li}
\email{xinqi.li@tju.edu.cn}
\affiliation{Center for Joint Quantum Studies and Department of Physics,
School of Science, Tianjin University, Tianjin 300072, China}

\author{Luting Xu}
\affiliation{Center for Joint Quantum Studies and Department of Physics,
School of Science, Tianjin University, Tianjin 300072, China}

\date{\today}

%% \maketitle
\begin{abstract}
The nonlocal nature of the Majorana zero modes implies an inherent teleportation
channel and unique transport signatures for Majorana identification.
In this work we make an effort to eliminate some inconsistencies
between the Bogoliubov-de Gennes equation based treatment
and the method using the associated regular fermion number states of occupation
within the `second quantization' framework.
We first consider a rather simple
`quantum dot--Majorana wire--quantum dot' system,
then a more experimentally relevant setup
by replacing the quantum dots with transport leads.
For the latter setup, based on the dynamical evolution of electron-hole excitations,
we develop a {\it single-particle-wavefunction} approach to quantum transport,
%{\color{red}\bf
which renders both the conventional quantum scattering theory
and the steady-state nonequilibrium Green's function formalism
as its stationary limit.     % } %==
%%
% {\color{red}\bf
Further, we revisit the issue of Majorana tunneling spectroscopy
and consider in particular the two-lead coupling setup.
We present comprehensive discussions with detailed comparisons,
and predict a zero-bias-limit conductance of $e^2/h$
(for symmetric coupling to the leads),
which is {\it a half} of the popular result of the zero-bias-peak,
or, the so-called {\it Majorana quantized conductance} ($2e^2/h$).    % } %===
The present work may arouse a need to reexamine some existing studies
and the proposed treatment is expected to be involved in analyzing
future experiments in this fast developing field.
\end{abstract}

% \pacs{03.65.Yz,03.65.Sq,31.15.xv,31.15.xg}

\maketitle

\section{Introduction}

In the past years the interests to the Majorana zero modes (MZMs)
in topological superconductors have been switched from a theoretical topic into
an active experimental field in condensed matter physics
\cite{Kita01,Bee13,Tew13,Sar15,Agu17a}.
In particular, proposals based on semiconductor nanowires \cite{Sar10,Opp10}
stimulated the initial experiment of Mourik {\it et al.} \cite{Kou12}
and subsequent experiments with transport features consistent with Majorana modes
\cite{Hei12,Fur12,Xu12,Fin13,Mar13,Mar16,Mar17,Kou18a}.
% ==  {Sar10,Opp10}  .. % == {Kou12,Hei12,Fur12 ..   Xu12,Fin13,Mar13}
%%
The nonlocal nature of the MZMs and the intrinsic non-Abelian braiding statistics,
both implying an immunity from the influence of local environmental noises, promise
a sound potential for topological quantum computation \cite{Kit03,Sar08,Sar15}.
To confirm the nonlocal nature of the MZMs,
beyond the local tunneling spectroscopy experiments mentioned above,
nonlocal transport signatures (including also nonlocal conductances
based on the three-terminal setup) have been investigated
\cite{Gla16,DS17d,Sch17,Mar16b,Mar17b,Mar18b,Mar18c,Kou18b,Flen19b},
% {Gla16,DS17d,Sch17}{Mar16b,Mar17b,Mar18b,Mar18c,Kou18b} {Flen19b}
together with evidences such as
the peculiar noise behaviors \cite{Dem07,Bee08,Law09,Li12,Li13,Shen12,Zoch13}
and the $4\pi$ periodic Majorana-Josephson currents
\cite{Kita01,Sar10,Opp10,Fu09,Cay17}.
In particular, some more recent studies were extensively
focused on distinguishing the nonlocal MZMs
from the topologically trivial Andreev bound states by transport measurements
\cite{DS17,Cay15,Cay16,Cay19,Cay19b,Agu17,Agu18,Agu19}.

Closely related to the nonlocal nature of the MZMs,
the so-called {\it teleportation} issue emerges as
the existence of a dramatic ultrafast electron transfer channel
\cite{Sem06ab,Lee08,Fu10,Li14b}.
Most strikingly, since the two MZMs at the ends of the quantum nanowire
can be located far away, the teleportation channel is somehow
indicating certain type of `superluminal' phenomenon \cite{Sem06ab,Lee08,Li14b}.
In particular, since this channel is usually mixed with the Andreev
process of electron-pair splitting,
in Ref.\ \cite{Fu10}, a truncated teleportation Hamiltonian
was derived by considering the nanowire in contact with
a floating mesoscopic superconductor, instead of the grounded one as usual.
There, the strong charging energy of the mesoscopic superconductor
rules out the Andreev process,
making thus only the teleportation channel survived.

The ability allowing ultrafast charge transfer through the teleportation channel
is rather transparent using the low-energy effective Hamiltonian
and within the framework of `second quantization', which simply manifests
the MZMs associated regular fermion state occupied or not,
i.e., the number state $|1\ra$ or $|0\ra$.
However, as we will show in this work, the conventional treatment
based on the well known Bogoliubov-de Gennes (BdG) equation
will encounter difficulty to restore this basic feature.
In this work we propose a solving method to eliminate the inconsistency
between these two types of treatments.
We notice that the standard BdG treatment has been widely involved in literature
\cite{Flen19b,Dem07,Bee08,Law09,Vu18}.
The present work may arouse a need to reconsider some transport signatures
associated with the Majorana nonlocal nature and teleportation channel.

We structure the paper as follows.
We first consider in Sec.\ II a rather simple setup
following Refs.\ \cite{Lee08,Fu10},
say, a `quantum dot--Majorana wire--quantum dot' system (see Fig.\ 1),
then in Sec.\ III the setup by replacing the dots with transport leads.
For the former setup, we focus on the issue of `teleportation',
and particularly propose a scheme to eliminate the inconsistency
between the Bogoliubov-de Gennes equation based treatment
and the method within the `second quantization' framework,
using the regular fermion number states of occupation.
For the latter setup, we first propose a {\it single-particle-wavefunction}
quantum transport approach,
% {\color{red}\bf
which renders both the conventional quantum scattering theory
and the steady-state nonequilibrium Green's function formalism
as its long time stationary limit.
Then, we revisit the Majorana tunneling spectroscopy
with comprehensive discussions and make a new prediction.   % }%==
Finally, we summarize the work in Sec.\ IV.

\begin{figure}[h]
\includegraphics[scale=0.85]{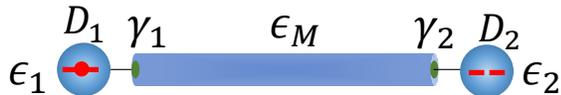}
\caption{
Schematic drawing for the setup of a Majorana quantum wire
coupled to two quantum dots.
The single electron is assumed initially in the left dot
and the subsequent evolution is expected to display
a `teleportation' type quantum oscillations
between the remotely distant dots.  }
%示意图：Majorana量子线左右各耦合一个量子点，初始态，量子点1 占据，量子点2不占据。
%中心的Majorana 量子线为低能有效模型，左右两端各有一个Majorana 束缚态$\gamma_1,\gamma_2$，
%量子线的哈密顿量为$H_M=i\gamma_1\gamma_2$.    }
%  \label{fig1}
\end{figure}

\section{Revisit the issue of Majorana teleportation}

\subsection{Low-Energy Effective Model and \\
   Number-State Treatment}

About the issue of `teleportation', let us consider first the simplest
`quantum dot--Majorana wire--quantum dot' setup (see Fig.\ 1),
following Refs.\ \cite{Lee08,Fu10},
to analyze the quantum transfer and oscillation
of an electron through a quantum wire
which accommodates a pair of Majorana bound states (MBSs).
The setup of Fig.\ 1 can be described by the following
effective low-energy Hamiltonian
\beq\label{Ham1}
    H=i\frac{\eps_M}{2}\gam_1\gam_2 +\sum_{j=1,2}
      \big[ \eps_j\ddg_j d_j+\lam_j(\ddg_j-d_j)\gam_j \big] \,.
\eeq
Here $\gam_1$ and $\gam_2$ are the Majorana operators
for the two MBSs at the ends of the quantum wire.
The two MBSs interact with each other by an energy $\eps_M$.
$d_1(\ddg_1)$ and $d_2(\ddg_2)$ are the annihilation (creation)
operators of the two single-level quantum dots,
while $\lam_1$ and $\lam_2$ are their coupling amplitudes to the MBSs.
The Majorana operators are related to the regular fermion
through the transformation of
$\gam_1=i(f-\fdg)$ and $\gam_2=f+\fdg$.
After an additional local gauge transformation,
$d_1\to id_1$, we reexpress Eq.\ (\ref{Ham1}) as
\bea\label{tunn-H-1}
H &=& \epsilon_M (f^{\dagger} f-\frac{1}{2})
+\sum_{j=1,2} [ \epsilon_j d^{\dagger}_j d_j
+\lambda_j (d^{\dagger}_j f+f^{\dagger} d_j) ]  \nl
&&-\lam_1(\ddg_1\fdg+f d_1)+\lam_2(\ddg_2\fdg+f d_2) \,.
\eea
It should be noticed that the tunneling terms in
this Hamiltonian only conserve charge modulo $2e$.
This reflects the fact that a pair of electrons
can be extracted from the superconductor condensate
and can be absorbed by the condensate vice versa.

Let us consider the transfer of an electron
between the two quantum dots, which is assumed
initially in the left quantum dot.
In particular, we consider the weak interaction limit $\eps_M\to 0$,
in order to reveal more drastically the teleportation behavior.
For simplicity, we assume $\lam_1=\lam_2=\lam$ and $\eps_1=\eps_2=0$.
Using the regular fermion number-state representation, i.e., $\ket{n_1, n_f, n_2}$,
where $n_{1(2)}$ and $n_f$ denote respectively the electron numbers
(``0" or ``1") in the left (right) dot and the central MZMs,
we have eight basis states:
$\ket{100}, \ket{010},\ket{001},\ket{111}$
with odd parity (electron numbers);
and $\ket{110},\ket{101},\ket{011},\ket{000}$ with even parity.
Associated with the specific initial condition, we only have
the odd-parity states involved in the state evolution.

Starting with the initial state $|100\ra$, the state evolution within the
odd-parity subspace can be carried out straightforwardly \cite{Lee08}.
Specifically, we are interested in the probability of electron
appearing in the right dot, which has two components \cite{Lee08}
\bea\label{telep}
P^{(1)}_2(\tau)&=& |\la 001 | e^{-iH \tau}\ket{100}|^2
=\sin^4(\lam \tau) \,, \nl
P^{(2)}_2(\tau)&=& |\la 111 | e^{-iH \tau}\ket{100}|^2
=\frac{1}{4}\sin^2(2\lam \tau)  \,.
\eea
Of great interest is the result of $P^{(1)}_2(\tau)$, which implies that,
even in the limit of $\eps_M\to 0$ (very long quantum wire),
the electron in the left dot can transmit through the quantum wire
and reappear in the right dot on a finite (short) timescale.
This is the remarkable `teleportation' phenomenon
discussed in Refs.\ \cite{Sem06ab,Lee08,Fu10} which,
surprisingly, holds a {\it superluminal} feature.

% {\color{red}\bf
However, the result of $P^{(2)}_2(\tau)$ is associated with the Andreev process,
i.e., splitting of a Cooper pair from the condensate of the superconductor.
To be more specific, let us consider the initial state $|100\ra$.
The state $|111\ra$ can be generated from $|100\ra$
by the local Andreev process at the right-hand-side,
which is described by the effective tunneling term
$d^{\dagger}_2 f^{\dagger}$ in \Eq{tunn-H-1}.
Obviously, this is not the event of teleportation of interest,
since the electron appearing in the right dot ($D_2$)
is not the one initially prepared in the left dot ($D_1$).
In order to single out the teleportation channel
from the Andreev process, it would be highly desirable
if we can suppress the terms
($\lambda_j d^{\dagger}_j f^{\dagger} +{\rm h.c.}$) in \Eq{tunn-H-1}.

Indeed, it was proposed in Ref.\ \cite{Fu10} that a nanowire is in proximity
contact with a mesoscopic {\it floating} superconductor
with strong charging energy $E_C$.
Under such assumptions, it was derived by an elegant and precise treatment that
the tunnel coupling is truncated to the following Hamiltonian
of tunneling through a single resonant level \cite{Fu10}
\bea\label{Fu-H}
    H &=&\eps_M (\fdg f-\frac{1}{2})
    +\sum_{j=1,2}\big[ \eps_j\ddg_j d_j+\lam_j(\ddg_j f+\fdg d_j)\big] \,. \nl
\eea
Comparing this result with the tunneling Hamiltonian in \Eq{tunn-H-1},
we find that the Andreev process terms have been ruled out and
that the only survived charge transfer channel is the {\it real}
transmission through the nonlocal Majorana states.
This is the true teleportation channel of our interest.

As an additional remark,
it should be noted that the suppression of the Andreev process terms
does not mean the superconducting pairing term destroyed.
Actually, the superconducting pairing term has been taken into account
when diagonalizing the superconductor Hamiltonian, which is responsible to
the formation of both the ground state condensate and the quasiparticle states
(including the Majorana $f$ quasiparticle).
The tunnel coupling Hamiltonians in both \Eqs{tunn-H-1} and (\ref{Fu-H})
are an effective low-energy description.      %  } %===

After suppressing the Andreev process, the transfer dynamics only involves states
$\ket{100}$, $\ket{010}$, and $\ket{001}$.
The time dependent state can be therefore expressed as
$|\Psi(\tau)\ra = a(\tau)\ket{100}+b(\tau)\ket{010}+c(\tau)\ket{001}$.
Also, we consider the simplest case by assuming
$\epsilon_M=\epsilon_1=\epsilon_2=0$ and $\lambda_1=\lambda_2=\lambda$.
Solving the Schr\"odinger equation based on the Hamiltonian \Eq{Fu-H} yields
\bea\label{abc-solution}
a(\tau)&=&\frac{1}{2} \left[ 1+\cos(\sqrt{2}\lambda\tau) \right] \,,  \nl
b(\tau)&=&-\frac{i}{\sqrt{2}} \sin(\sqrt{2}\lambda\tau) \,, \nl
c(\tau)&=&\frac{1}{2} \left[ -1+\cos(\sqrt{2}\lambda\tau) \right] \,.
\eea
This solution was obtained with the initial condition $|\Psi(0)\ra = \ket{100}$.
Therefore, the occupation probability of the right dot,
$P_2(\tau)=|c(\tau)|^2=\sin^4(\lambda\tau/\sqrt{2})$,
reveals a {\it real} teleportation feature
as discussed above based on $P_2^{(1)}(\tau)$ in \Eq{telep}.
In Fig.\ 2(a), using the above analytic solution,
we plot the occupation probabilities
of the two dots (by the black and red lines).
The displayed simple quantum oscillations are indeed remarkable,
viewing that the two dots are coupled through a very long quantum wire.

\subsection{Bogoliubov-de Gennes Equation Based Simulation}

We now turn to a lattice-model-based simulation for the above transfer dynamics
using the BdG equation
and the well known Kitaev model for the topological quantum wire \cite{Kita01}
\bea
H_W &=& \sum_j \left[
-\mu c^{\dg}_{j}c_j - \frac{t}{2}  (c^{\dg}_{j}c_{j+1}+{\rm h.c.}) \right] \nl
&& + \frac{\Delta}{2} \sum_j (c_{j}c_{j+1}+{\rm h.c.})  \,.
\eea
In this spinless $p$-wave superconductor model,
$\mu$ is the chemical potential,
$\Delta$ is the superconducting order parameter,
and $t$ is the hopping energy between the nearest neighbor sites
with $c^{\dagger}_j$ ($c_j$) the associated electron creation (annihilation) operators.
The specific choice of $\frac{t}{2}$ and $\frac{\Delta}{2}$
is for a convenience such that the energy gap parameter
of the quasiparticle excitations is $\Delta$ (rather than $2\Delta$).
The total Hamiltonian of the setup shown in Fig.\ 1 reads $H=H_W+H_D+H'$,
with $H_D=\sum_{j=1,2}\epsilon_j d^{\dagger}_jd_j$
and the coupling between the dots and the quantum wire given by
\bea
H'=(t_L d_1c^{\dagger}_1 +  t_R d_2c^{\dagger}_N)+{\rm h.c.} \,,
\eea
with $t_L$ and $t_R$ the coupling energies.

In order to introduce the representation of electron and hole states,
we use the Nambu spinor
$\hat{\Psi}=(c_1,\cdots,c_N,c^{\dg}_1,\cdots,c^{\dg}_N)^T$
and rewrite the Hamiltonian of the quantum wire as
$H_W=\frac{1}{2}\hat{\Psi}^{\dg}\tilde{H}_W\hat{\Psi}$,
which yields thus the BdG Hamiltonian matrix
\begin{eqnarray}
\tilde{H}_W =  \left(
\begin{array}{cc}
T  &  \Omega  \\
-\Omega  &  -T
\end{array}
\right) \,,
\end{eqnarray}
where the block elements are given by
\begin{eqnarray}
T= \left(
\begin{array}{ccccc}
-\mu & -t/2 & 0 & \cdots & \cdots  \\
-t/2 & -\mu & -t/2 & 0 & \cdots  \\
0 & -t/2 & -\mu & -t/2 & \cdots   \\
\cdot & \cdot & \cdot & \cdot & \cdot  \\
\cdot & \cdot & \cdot & \cdot & \cdot
\end{array}
\right)  \,,
\end{eqnarray}
and
\begin{eqnarray}
\Omega= \frac{1}{2}\left(
\begin{array}{ccccc}
0 & \Delta & 0 & \cdots & \cdots  \\
-\Delta & 0 & \Delta & 0 & \cdots  \\
0 & -\Delta & 0 & \Delta & \cdots   \\
\cdot & \cdot & \cdot & \cdot & \cdot  \\
\cdot & \cdot & \cdot & \cdot & \cdot
\end{array}
\right) \,.
\end{eqnarray}
More physically, the above BdG Hamiltonian matrix can be understood
as being constructed under the single-particle basis
$\{ |e_1\ra,\cdots,|e_N\ra;~ |h_1\ra,\cdots,|h_N\ra \}$,
where $|e_j\ra$ and $|h_j\ra$ describe,
respectively, the electron and hole states on the $j$th site.

Further, let us consider the entire `Dot-Wire-Dot' system.
Using the joint electron and hole basis,
the complete states of the quantm dots
should include both $|D_j\ra$ and $|H_j\ra$,
with $j=1,2$ labeling the quantum dots while `$D$' and `$H$'
describing the electron and hole states, respectively.
Accordingly, the Hamiltonian should include couplings
of $|D_1\ra$ with $|e_1\ra$ and $|D_2\ra$ with $|e_N\ra$ for electrons,
and $|H_1\ra$ with $|h_1\ra$ and $|H_2\ra$ with $|h_N\ra$ for holes.
It is well known that
the hole couplings are employed to describe the Andreev process.
For instance, in the simplified description of the low-energy excitations,
the transition $|1,0,0\ra \Rightarrow|1,1,1\ra$
corresponds to annihilating the hole state $|H_2\ra$
(owing to the transfer of $|H_2\ra$ to $|h_N\ra$),
and at the same time exciting the `$f$' quasi-particle of the MZMs
(via the $|h_N\ra$ excitation).
Similarly, the transition $|1,1,1\ra \Rightarrow|0,0,1\ra$
is mediated by the hole transfer from $|h_1\ra$ of the wire
to $|H_1\ra$ of the left dot.

To make a close comparison between the effective low-energy model result
and the Kitaev lattice model based simulation, we restrict our analysis
to the transfer dynamics associated with
the truncated `teleportation' Hamiltonian, \Eq{Fu-H},
where only the teleportation channel is left
while the Andreev process is suppressed.
Then, in the absence of hole couplings between the dots and the quantum wire,
the coupling Hamiltonian reads
\bea\label{tunnel-H2}
H'=( t_L |e_1\ra\la D_1| +  t_R |e_N\ra\la D_2|)+ {\rm h.c.}  \,.
\eea
Again, let us consider the evolution starting with $|\Psi(0)\ra=|D_1\ra$,
i.e., initially the electron in the left dot.
The transfer dynamics is described by
\bea\label{WF-1}
&&|\Psi(\tau)\ra
= \alpha_1(\tau) |D_1\ra + \alpha_2(\tau) |D_2\ra \nl
&& ~~~~~~ + \sum^N_{j=1}[u_j(\tau)|e_j\ra+v_j(\tau)|h_j\ra] \,,
\eea
where the superposition coefficients can be solved
from the time-dependent Schr\"odinger equation,
$i\hbar\frac{\partial}{\partial \tau}|\Psi(\tau)\ra = H |\Psi(\tau)\ra$,
by casting the Hamiltonian into the BdG-type matrix form,
using the joint electron and hole basis.

\begin{figure}[h]
  %\centering
  \includegraphics[scale=0.8]{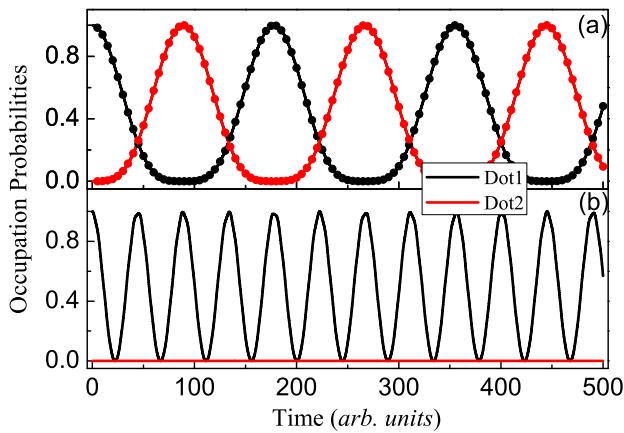}
  \caption{
Quantum oscillations of an electron between two remote quantum dots,
mediated by the nonlocal MZMs.
(a)
Plots of the analytic solution \Eq{abc-solution} (black and red lines,
based on the number-states treatment of the low-energy effective model),
compared with the results from the lattice model based simulation
using the tunneling Hamiltonian \Eq{tunnel-H3} (black and red dots).
Through the whole work we adopt an arbitrary system of units
by setting the hopping energy in the Kitaev lattice model $t=1$.
Other parameters in the lattice model:
$\epsilon_1=\epsilon_2=0$, $t_L=t_R=0.05$, $\mu=0$ and $\Delta=1.0$.
The corresponding parameters of the reduced low-energy effective model:
$\epsilon_M=0$ and $\lambda_1=\lambda_2=\lambda=0.025$.
(b)
Results based on the Kitaev lattice model and
using the tunneling Hamiltonian \Eq{tunnel-H2},
which involves both the positive and negative energy eigenstates in the dynamics.
Parameters are the same as in (a). }
\end{figure}

In Fig.\ 2(b) we show the results from numerically solving \Eq{WF-1}.
To compare with the results displayed in Fig.\ 2(a),
we plot the probabilities $P_1(\tau)=|\alpha_1(\tau)|^2$
and $P_2(\tau)=|\alpha_2(\tau)|^2$ by the black and red lines, respectively.
Most surprisingly, in Fig.\ 2(b), we find no occupation of the right dot
with the increase of time, which simply means
no charge transfer mediated by the MZMs.
We only find quantum oscillations between the left dot and the quantum wire,
but with a period differing from that in Fig.\ 2(a),
despite that in both plots we have used identical coupling strengths.
We may identify the reasons for both results as follows.

By diagonalizing the BdG Hamiltonian $\tilde{H}_W$ of the quantum wire,
one obtains two sets of eigenstates, say, $|E_n\ra$ and $|-E_n\ra$ with
$n=1,2,\cdots,N$, corresponding to the positive and negative eigen-energies.
In particular, in the topological regime, the lowest energy states
$|E_1\ra$ and $|-E_1\ra$ are sub-gap states with $E_1\to 0$
and the wavefunctions distribute at the ends of the quantum wire.
The MBSs at the ends of the wire are obtained from, respectively,
$|\gamma_1\ra=(|E_1\ra+|-E_1\ra)/2$
and $|\gamma_2\ra=(|E_1\ra-|-E_1\ra)/2i$.
From the tunnel Hamiltonian \Eq{tunnel-H2}, the charge transfer $|D_1\ra\to |e_1\ra$
will generate a quantum superposition of $|E_1\ra$ and $|-E_1\ra$
in the quantum wire, especially with equal weights as $E_1\to 0$.
Owing to the requirement of energy conservation,
the higher eigen-energy states will not be excited (populated)
after a timescale longer than $\hbar/t_L$.
As a consequence of this superposition of $|E_1\ra$ and $|-E_1\ra$,
the electron and hole excitations are largely located at
the left side of the wire, leading thus to {\it no charge transfer}
to the right side of the wire and to the right side quantum dot.

The simultaneous coupling of $|D_1\ra$ to the zero-energy states
$|E_1\ra$ and $|-E_1\ra$ of the quantum wire
is also the reason for the different periods of oscillations in Fig.\ 2(b) and (a).

We understand then that the main difference of the coupling Hamiltonian \Eq{tunnel-H2}
from the `number'-states treatment using the low-energy effective model
is the {\it redundant coupling} of the dot electron
to the negative-energy eigenstates of the superconducting quantum wire.
Indeed, the negative-energy eigenstates are the dual counterparts
of the Bogoliubov quasi-particles (the positive-energy eigenstates).
Before diagonalizing the Hamiltonian of the superconductor,
introducing holes (with negative energies) is unavoidable,
in order to `mix' the electron and hole components
to form the Bogoliubov quasi-particles
(physically, owing to the many body electron-electron scattering
and the existence of the superconducting condensate).
However, after the diagonalization,
the negative-energy eigenstates are redundant.
A negative-energy eigenstate simply means
the result of removing an existing quasi-particle
(which has positive energy).
Moreover, the corresponding Bogoliubov `creation' operators
of the negative-energy eigenstates will, importantly,
{\it annihilate} the ground state of the superconductor.
In other words, {\it the negative-energy eigenstates cannot be created
from the ground state of the superconductor}.
Therefore, if we explicitly introduce the creation of
Bogoliubov positive-energy quasiparticles (from the ground state)
and annihilation of the existing ones,
the negative-energy eigenstates are redundant,
which should not appear in the tunnel coupling Hamiltonian.

For the specific setup under consideration,
the tunnel coupling Hamiltonian should thus be modified as
\bea\label{tunnel-H3}
H'=( t_L |\tilde{e}_1\ra\la D_1|
+  t_R |\tilde{e}_N\ra\la D_2|)+ {\rm h.c.}  \,,
\eea
where the two projected states are defined through
\bea
|\tilde{e}_1\ra &=& \hat{P}|e_1\ra \,, \nl
|\tilde{e}_N\ra &=& \hat{P}|e_N\ra \,,
\eea
while the projection operator is defined by
\bea\label{Proj}
\hat{P}=\sum^N_{E_n>0,~n=1}|E_n\ra\la E_n| \,.
\eea
Very importantly, the above tunneling Hamiltonian properly accounts for
the creation and annihilation of the Bogoliubov quasiparticles
(with positive energies),
which are the {\it real existence} in superconductors.
Here, owing to the suppression of the Andreev process,
the hole states of the quantum dots do not appear
in the tunnel coupling to the Bogoliubov quasiparticles.
Otherwise, in the presence of Andreev process, as we will see later,
the hole states of the transport leads will
participate in the coupling to the Bogoliubov quasiparticles.

Based on the tunnel Hamiltonian \Eq{tunnel-H3}, we re-simulate
the electron transfer dynamics and obtain results
shown in Fig.\ 2(a) by the symbols of black and red dots.
In contrast to what we observed in Fig.\ 2(b),
here the desired quantum oscillations are recovered
in precise agreement with the number-state treatment
based on the low-energy effective model.
%%
% {\color{red}\bf
We should mention that
this full agreement is achieved in the regime of {\it weak coupling}
between the dots and the quantum wire, which guarantees
the dominant coupling of the quantum dots being to the MZMs,
but not to the Bogoliubov quasiparticle states above the superconducting gap.
If we consider a {\it strong coupling} regime,
occupation of the quasiparticle states above the gap
will result in some irregularities instead of the ideal quantum oscillations.  % } %===

\subsection{On the Teleportation Issue}

\begin{figure}[h]
  \includegraphics[scale=0.85]{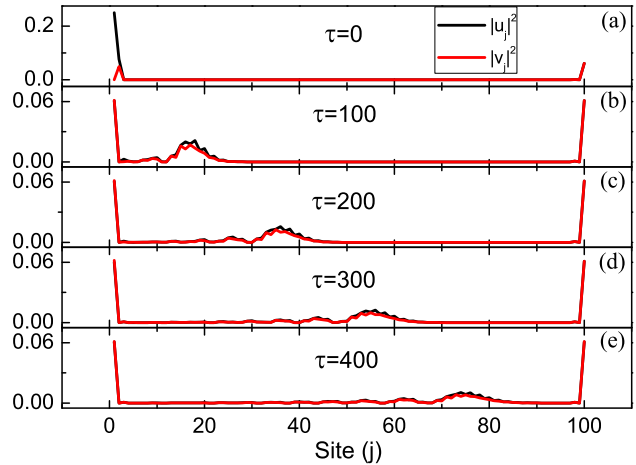}
  \caption{
Wavepacket propagation of the electron-hole excitations
based on simulation of the isolated Kitaev's lattice wire.
We adopt an arbitrary system of units
by setting the hopping energy $t=1$ (and $\hbar=1$),
and assume the other parameters $\mu=0$ and $\Delta=0.8$.
(a)
Initial distribution of the electron-hole excitations
after projecting the lattice state $|e_1\ra$ onto
the Hilbert space of the Bogoliubov quasiparticle states,
which corresponds to the action of $c_1^{\dagger}$
on the superconductor ground state $|G\ra$.
(b)-(e)
Propagation of the electron-hole excitations along the lattice wire
with the increase of time ($\tau$).   }
\end{figure}

%{\color{red}\bf
%%
Taking the lattice model, let us first simulate
the `microscopic' dynamics of the electron-hole excitations in the quantum wire.
Without loss of the main physics, as a simpler and clearer illustration,
we may consider an {\it isolated} quantum wire
with an initial excitation of $|e_1\ra$.
This corresponds to the electron in the left dot ($D_1$)
entering the wire via the first site,
i.e., $|\tilde{e}_1\ra=c_1^{\dagger}|G\ra=\hat{P}|e_1\ra$,
where $|G\ra$ is the ground state of the superconductor wire.
Note that, owing to the property (requirement) of the ground state,
the real {\it physical state} is $|\tilde{e}_1\ra$ but not $|e_1\ra$.

Specifically, let us assume $\mu=0$ and $\Delta=0.8$
(note that we always set $t=1$ through the whole work).
We find that, from the initial `lattice state' $|e_1\ra$,
the projection probability of getting the Majorana state $|E_1\ra$
is, $p_1=|\la E_1|e_1\ra|^2\simeq 0.247$.
In the ideal case of $\Delta=t$, this probability is $p_1=0.25$.
In Fig.\ 3(a), we show the initial electron-hole excitations
from the (unnormalized) projected state $|\tilde{e}_1\ra=\hat{P}|e_1\ra$.
The result displayed by the red curve in Fig.\ 3(a)
is the distribution of the hole components on the lattice sites,
which largely characterizes the distribution of the electron-hole excitations
in the Majorana state $|E_1\ra$, by noting that
the weights of the electron and hole components are equal
i.e., $|u_j|^2=|v_j|^2$ on every lattice site, for $E_1\simeq 0$.
However, on the left side,
owing to the quantum superposition with the high energy states
and thus the quantum interference,
the hole distribution has some distortion
compared to the right side, where this same effect is negligibly weak.
The black curve in Fig.\ 3(a)
more heavily involves the electron-component contribution
of the high energy states above the gap,
with their superposition resulting in the localized distribution in space.
In Fig.\ 3(b)-(e),
we show the wavepacket propagation of the electron-hole excitations.
It should be noted that the propagation is largely from
the electron-hole excitations of the high energy states.
The two-side edge distributions associated with zero-energy Majorana mode
are almost unaffected from the evolution.

The picture revealed in Fig.\ 3 indicates that the charge transfer
mediated by the Majorana state is {\it not} via the wavepacket
propagation of the electron-hole excitations along the wire.
Once the external electron enters the wire,
the two-side edge excitations associated with the MZMs
are generated instantaneously, thus holding
the `teleportation' ability to mediate charge transfer.
We may argue this extremely puzzling issue with a few remarks in order as follows.

%\begin{enumerate}
%\item

{\it (i)}
We notice that in Ref.\ \cite{Lee08}, in order to rule out
the difficulty of arriving to a `superluminal' conclusion,
it was argued that since a classical exchange
of information (the result of the coincident measurements) is necessary,
there is no superluminal transfer
of information in the observation of the teleportation effect.
However, as pointed out in Ref.\ \cite{Li14b},
in principle one can confirm the `teleportation event' in a later stage
from the {\it coincident measurement data} of two remote detectors.
Obviously, this confirmation for the {\it existing objective event}
does not need any communication of classical information.

%\item
%(ii) \\
{\it (ii)}
Unlike the argument in Ref.\ \cite{Lee08},
we may provide an alternative understanding.
Since the Majorana state is a quasiparticle excitation in the presence
of many other electrons (condensate of the superconductor ground state),
we cannot conclude that the electron appeared in the right
quantum dot is the one initially in the left dot.
%% 组态  重组
Indeed, consider the action of $c_1^{\dagger}$ on
the superconductor ground state $|G\ra$.
viewing that $|G\ra$ is a condensate of many electrons
(superposition of occupied and unoccupied electron pairs),
the action of $c_1^{\dagger}$ would induce a `reformation' of
the whole correlated many electron condensate.
From the `re-organized' condensate, the quasiparticle excitation
can be separated with respect to the ground state.
In particular, the Majorana state among the quasipaiticle excitations
holds the nonlocal nature, as a superposition of the electron-hole excitations
at the two ends of the quantum wire.

Obviously, the electron-hole components at the right side
are {\it not} from the left side through any quantum transfer process.
The new particle is formed as a result of `re-organization'
of the many-particle condensate.
This re-organization process, which thus allows us to extract electron
from the right side,
may resemble in some sense the current formation in a conducting wire under electric field,
where the current forming at a remote place is not from the electrons
of the initial place we performed electric disturbance.
The speed of current formation in the conducting wire is not superluminal.
Similarly, the `reorganization' of the many electron condensate mentioned above
cannot be superluminal.
Therefore, the `superluminal' feature of Majorana teleportation is a result
that the `re-organization' process of the many electron condensate
did not enter a dynamical description.

%\item
{\it (iii)}
The action of the {\it local} operator of $c_1^{\dagger}$ on the ground state,
which causes the {\it nonlocal} excitation of the Majorana state,
can be also understood from the perspective of quantum measurement.
More specifically,
in terms of the POVM (positive-operator-value-measure) formalism,
let us consider $\tilde{\rho}= M\rho M^{\dagger}/||\bullet||$,
with the Kraus measurement operator $M=c_1^{\dagger}$,
the density matrix of the ground state $\rho=|G\ra \la G|$
and $||\bullet||$ denoting the normalization factor.
We know that $c_1^{\dagger}$ can be decomposed into a superposition
of the Bogoliubov operators, both the creation and annihilation operators.
However, the action of the annihilation operators on the ground state
would vanish the result.
The state survived from this action is a superposition
of the quasiparticle states generated by the creation operators.
Among them, the particular Majorana state is highly nonlocal.
Actually, the `measurement' process described by the POVM projection should
correspond to the `re-organization' of the many electron condensate.

%\end{enumerate}

To summarize,
the Majorana-nonlocality-induced teleportation looks like a superluminal phenomenon,
but in reality it cannot be, if we take into account the re-organization process
of the many electron condensate and/or the measurement process discussed above.     % } %=====

\section{Transport through Majorana quantum wires}

\subsection{Preliminary Consideration}

As a more realistic configuration,
let us consider to connect the quantum wire
with two transport leads, instead of the quantum dots.
The transport leads can be described by the interaction-free Hamiltonian
\bea
H_{\rm leads}=\sum_l \epsilon_l a^{\dagger}_l a_l
+ \sum_r \epsilon_r b^{\dagger}_r b_r  \,,
\eea
and the coupling of the quantum wire to the leads
is described by the tunnel Hamiltonian
\bea
H'= \left( \sum_l t_l  c^{\dagger}_1 a_l
  + \sum_r t_r   c^{\dagger}_N b_r \right) + {\rm h.c.}  \,.
\eea
To display the Andreev process in a transparent manner,
let us introduce the electron and hole basis
$\{|e_j\ra, |h_j\ra ~|~ j=1,2,\cdots,N  \}$ for the Kitaev quantum wire,
and similarly $\{|e_l\ra, |h_l\ra \}$
and $\{|e_r\ra, |h_r\ra \}$ for the left and right leads.
Using these basis states, the tunnel Hamiltonian can be rewritten as
\bea
&& H'= \left[ \sum_l t_l (|e_1\ra\la e_l|-|h_1\ra\la h_l|)  \right.  \nl
  && ~~~ + \left. \sum_r t_r (|e_N\ra\la e_r|-|h_N\ra\la h_r|) \right] +{\rm h.c.} \,.
\eea
In particular, the tunnel coupling between the hole states in this form
is explicitly used to describe the Andreev process.
However, based on the lesson learned in the `Dot-Wire-Dot' setup,
we propose to modify the tunnel Hamiltonian as
\bea\label{tunnel-H4}
&& H'= \left[
\sum_l t_l (|\tilde{e}_1\ra\la e_l|-|\tilde{h}_1\ra\la h_l|)  \right.  \nl
&& ~~~ + \left.
\sum_r t_r (|\tilde{e}_N\ra\la e_r|-|\tilde{h}_N\ra\la h_r|) \right] +{\rm h.c.} \,,
\eea
where the lattice edge site states (for both electrons and holes)
are projected onto the subspace of the Bogoliubov quasiparticle states,
through the projector $\hat{P}$ introduced previously by \Eq{Proj}.

\subsection{Single Particle Wavefunction Approach}

For mesoscopic quantum transports, there exist well known approaches
such as the nonequilibrium Green's function (nGF) method \cite{Jau96,Datt95}
and the S-matrix quantum scattering theory \cite{Datt95,Gla02}
which are particularly suitable,
in the absence of many-body interactions, to study transport
through a large system modeled by the tight-binding lattice model
and with superconductors involved (either as the leads or a central device).
Another less-developed method, say,
the single particle wavefunction (SPWF) approach \cite{SG91,Li09,Li12b,SG16},
is an alternative but attractive choice.
This method, directly based on the time-dependent Schr\"odinger equation,
was developed in the context of transport through small systems
such as quantum dots
and has been applied skillfully to study some interesting problems \cite{SG16}.
%% //
Below we extend it to study quantum transports through large lattice systems,
especially in the presence of superconductors which may result in rich physics
such as Andreev reflections and phenomena related to the MZMs.
Importantly, this method can be regarded as
an extension of the S-matrix scattering theory,
i.e., from {\it stationary} to {\it transient} versions.
For instance, this method should be very useful
to study the possible transport probe
of non-adiabatic transitions during Majorana braiding
in the context of topological quantum computations.

The basic idea of the SPWF method is
keeping track of the quantum evolution of an electron initially in the source lead,
based on the time-dependent Schr\"odinger equation,
and computing various transition rates such as
the transmission rate to the drain lead,
or Andreev-reflection rate back to the source lead as a hole.
For the problem under study,
we denote the initial state as $|\Psi(0)\ra=|e_{\bar{l}}\ra$.
The subsequent evolution will result in a superposition
of all basis states of the leads and the central device, expressed as
\bea
&&|\Psi\ra = |\Psi_w\ra + |\Psi_{\rm leads}\ra  \nl
&& = \sum^N_{j=1} (u_j|e_j\ra+v_j|h_j\ra)
+ \sum_l (\alpha_l|e_l\ra+\tilde{\alpha}_l|h_l\ra)  \nl
&&
~~ + \sum_r (\beta_r|e_r\ra+\tilde{\beta}_r|h_r\ra) \,.
\eea
Based on the time dependent Schr\"odinger equation,
$i |\dot{\Psi}\ra = H |\Psi\ra$, we have
\bea
i\,\dot{u}_j &=& (\bullet)
+ \sum_l t_{l} \alpha_l \la e_j|\tilde{e}_1\ra
+ \sum_l (-t_{l}) \tilde{\alpha}_l \la e_j|\tilde{h}_1\ra   \nl
&& ~~ + \sum_r t_{r} \beta_r \la e_j|\tilde{e}_N\ra
+ \sum_r (-t_{r}) \tilde{\beta}_r \la e_j|\tilde{h}_N\ra   \nl
i\,\dot{v}_j &=& (\bullet)
+ \sum_l t_{l} \alpha_l \la h_j|\tilde{e}_1\ra
+ \sum_l (-t_{l}) \tilde{\alpha}_l \la h_j|\tilde{h}_1\ra   \nl
&& ~~ + \sum_r t_{r} \beta_r \la h_j|\tilde{e}_N\ra
+ \sum_r (-t_{r}) \tilde{\beta}_r \la h_j|\tilde{h}_N\ra   \nl
i\,\dot{\alpha}_l &=& \epsilon_l \alpha_l
   + t^*_{l} \la \tilde{e}_1|\Psi_w\ra  \nl
i\,\dot{\tilde{\alpha}}_l &=& -\epsilon_l \tilde{\alpha}_l
- t^*_{l} \la \tilde{h}_1 |\Psi_w\ra  \nl
i\,\dot{\beta}_r &=& \epsilon_r \beta_r + t^*_{r} \la \tilde{e}_N|\Psi_w\ra  \nl
i\,\dot{\tilde{\beta}}_r &=& -\epsilon_r \tilde{\beta}_r
- t^*_{r } \la \tilde{h}_N|\Psi_w\ra
\eea
For the sake of brevity,
in the first two equations, we have used the symbol $(\bullet)$
to denote the terms for the central system (in the absence of coupling to leads).
Performing the Laplace and inverse-Laplace transformations,
after some algebras, we obtain
\bea
i\,\dot{u}_j &=& (\bullet) -i\frac{\Gamma_L}{2}
\, \left[ \la e_j|\tilde{e}_1 \ra \la \tilde{e}_1 |\Psi_w\ra
+ \la e_j|\tilde{h}_1 \ra \la \tilde{h}_1|\Psi_w\ra  \right] \nl
&&  -i\frac{\Gamma_R}{2}
\, \left[ \la e_j|\tilde{e}_N\ra \la \tilde{e}_N|\Psi_w\ra
+ \la e_j|\tilde{h}_N\ra \la \tilde{h}_N|\Psi_w\ra  \right]  \nl
&& + t_L e^{-iE_{\rm in}t} \la e_j|\tilde{e}_1\ra   \nl
i\,\dot{v}_j &=& (\bullet) -i\frac{\Gamma_L}{2}
\, \left[ \la h_j|\tilde{e}_1 \ra \la \tilde{e}_1 |\Psi_w\ra
+ \la h_j|\tilde{h}_1 \ra \la \tilde{h}_1|\Psi_w\ra  \right] \nl
&&  -i\frac{\Gamma_R}{2}
\, \left[ \la h_j|\tilde{e}_N\ra \la \tilde{e}_N|\Psi_w\ra
+ \la h_j|\tilde{h}_N\ra \la \tilde{h}_N|\Psi_w\ra  \right]  \nl
&& + t_L e^{-iE_{\rm in}t} \la h_j|\tilde{e}_1\ra
\eea
In a more compact form, the result can be reexpressed as
%\begin{subequations}
\begin{eqnarray} \label{result-1}
i \begin{bmatrix}
\dot{u}_1 \\ \dot{u}_2\\ \vdots \\\dot{u}_N  \\
\dot{v}_1 \\ \dot{v}_2\\ \vdots \\\dot{v}_N  \end{bmatrix}
=  (\bullet) + \left( \hat{P} \Sigma \hat{P}  \right)
\begin{bmatrix}
u_1 \\ u_2\\ \vdots \\ u_N  \\
v_1 \\ v_2\\ \vdots \\  v_N  \end{bmatrix}
+ t_L e^{-i E_{\rm in}t} \hat{P}
\begin{bmatrix}
1 \\ 0 \\ 0 \\ 0 \\ \vdots \\ 0 \\ 0 \\ 0 \\ 0  \end{bmatrix}
\end{eqnarray}
%\end{subequations}
where we introduce the self-energy operator as
\bea\label{result-2}
&&\Sigma = (-i\Gamma_L/2)\,(|e_1\ra\la e_1|+|h_1\ra\la h_1|) \nl
&& ~~ + (-i\Gamma_R/2)\,(|e_N\ra\la e_N|+|h_N\ra\la h_N|)  \,.
\eea
\Eq{result-1} describes the evolution dynamics of the electron-hole excitations,
in the presence of tunnel-couplings to the transport leads
which lead to the self-energy term, i.e., the second term
on the right-hand-side (r.h.s) of \Eq{result-1} together with \Eq{result-2}.
The third term on the r.h.s of \Eq{result-1} is resulted from the
tunnel-coupling which injects the initial electron into the quantum wire.
For both of the two terms, only the real (positive-energy) Bogoliubov
quasiparticle states participate in the tunneling process,
as imposed by the projection operator.
Again, we emphasize that the projection eliminates the {\it redundancy} (`double-use')
of the negative-energy eigenstates to be involved in the tunneling process.
Physically speaking, the negative-energy eigenstate simply means the consequence
of annihilating an existing positive-energy quasiparticle via, for instance,
the usual tunneling or the more dramatic Andreev process.
These two processes, by using only the positive-energy eigenstates,
have been already accounted for in the treatment of the tunnel couplings,
i.e., in \Eq{tunnel-H4}.
However, we may notice that \Eq{result-1} does not exclude
any possible presence of the negative-energy eigenstates
during the inside electron-hole excitation dynamics in the quantum wire.

It is clear that,
based on the time-dependent state $|\Psi_w(t)\ra$ given by \Eq{result-1},
one can straightforwardly compute the various current components
by finding first the {\it projected} occupation probabilities
of the terminal sites of the quantum wire
(for both the electron and hole components),
then multiplying the tunnel-coupling rates, which yields
\bea\label{result-3}
i_{\rm LR} &=& e\Gamma_R\, |\la e_N|\hat{P}|\Psi_w\ra|^2 \,,  \nl
i_{\rm A} &=&  e\Gamma_L\, |\la h_1|\hat{P}|\Psi_w\ra|^2  \,,  \nl
i_{\rm CA} &=&  e\Gamma_R\, |\la h_N|\hat{P}|\Psi_w\ra|^2  \,,
\eea
where $e$ is the electron charge. These are the single-incident-electron
(initially in $|e_{\bar{l}}\ra$) contributed current components
associated with, respectively,
the normal electron transmission from the left to right leads,
the local Andreev reflection at the left side,
and the cross Andreev reflection process.

\subsection{Connection with Other Approaches}

To express the results in a more general form,
let us denote the incident channel by $\alpha$,
the outgoing channel by $\beta$,
and the associated current by $i_{\alpha\beta}$.
The `total' current associated with the $(\alpha,\beta)$ channels
from the incident electrons within the unit energy interval around $E$
is simply given by $\rho_{\alpha}(E) i_{\alpha\beta}(E)$,
with $\rho_{\alpha}(E)$ the density-of-states at the incident energy.
In long time limit (stationary limit), comparing this result with the current
derived from the nonequilibrium Green's function (nGF) technique
\cite{Jau96,Datt95},
we can establish the following connection between the two approaches
\bea\label{i-T}
\rho_{\alpha}(E)\, i_{\alpha\beta}(E)=\frac{e}{h}\,{\cal T}_{\alpha\beta}(E) \,.
\eea
In this expression, $h$ is the Plank constant and ${\cal T}_{\alpha\beta}(E)$
is the transmission coefficient from the channel $\alpha$ to $\beta$
at the energy $E$, which can be used to compute the linear-response or differential
conductance by means of the well-known Landauer-B\"uttiker formula as
$G_{\alpha\beta}=(e^2/h){\cal T}_{\alpha\beta}$.
In this context, we like to mention that for the {\it two-electron} Andreev reflections,
the respective conductance is related to the hole-reflection coefficient as
$G_{\rm A}=(2e^2/h){\cal T}_{\rm A}$.
Within the nGF formalism, the transmission coefficient is given by \cite{Jau96,Datt95}
\bea\label{nGF-1}
{\cal T}_{\alpha\beta}(E)={\rm Tr}(\Gamma_{\alpha}G^r\Gamma_{\beta}G^a) \,,
\eea
where $G^{r(a)}$ is the retarded (advanced) Green's function
of the transport central system,
which includes the self-energies from the transport leads.
Notice that, even within the nGF formalism,
this result is valid only for transport through noninteracting systems.
Another connection is that this formula corresponds to the
S-matrix scattering approach \cite{Flen19b,Bee08,Law09,Gla02}
after summing all the final states of the scattering probability
under the restriction of energy conservation,
and for all the initial states at the energy $E$.

Applying the formula \Eq{nGF-1} to transport through a superconductor,
straightforwardly, we can obtain the coefficients of the electron transmission
(from left to right leads), the local Andreev reflection (in the left lead),
and the cross Andreev reflection, respectively, as \cite{Sun01,Sun09,Li14}
\bea\label{nGF-2}
{\cal T}_{\rm LR}(E)&=&{\rm Tr}\left(
\Gamma^e_L G^r_{ee} \Gamma^e_R G^a_{ee} \right)   \,, \nl
{\cal T}_{\rm A}(E)&=&{\rm Tr}\left(
\Gamma^e_L G^r_{eh} \Gamma^h_L G^a_{he} \right)   \,, \nl
{\cal T}_{\rm CA}(E)&=&{\rm Tr}\left(
\Gamma^e_L G^r_{eh} \Gamma^h_R G^a_{he} \right)   \,.
\eea
Here we have added explicitly the superscripts
`$e$' (for electrons) and `$h$' (for holes)
to the tunnel-coupling rates $\Gamma_{L}$ and $\Gamma_{R}$.
We have also expressed the Green's functions
in an explicit form of matrix sector
in the Nambu representation between the electron/hole states.

% {\color{red}\bf
The above results of \Eqs{i-T}-(\ref{nGF-2}) establish
a connection at steady-state transport limit
between the SPWF and nGF approaches,
based on the standard BdG treatment without projection onto
the space of positive-energy Bogoliubov quasiparticle states.
In order to account for the modified treatment with projection,
as a long-time stationary limit of \Eq{result-3},
we only need to modify the Green's functions in \Eq{nGF-2} as
$\widetilde{G}^{r(a)}=\hat{P}G^{r(a)}\hat{P}$,
together with the modified self-energies
$\widetilde{\Sigma}^{r(a)}=\hat{P}\Sigma^{r(a)}\hat{P}$,
as similarly done in \Eq{result-1} with the result of \Eq{result-2}.    % }

\begin{figure}[h]
 % \centering
  \includegraphics[scale=0.8]{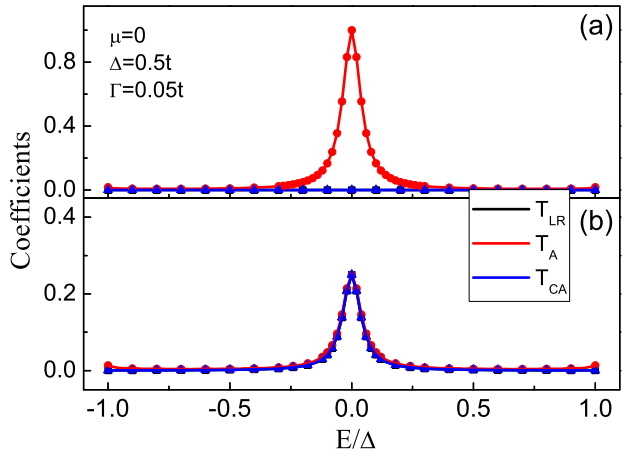}
  \caption{
% {\color{red}\bf
Kitaev's lattice model based simulation for coefficients of transmission
from the left to right leads (${\cal T}_{\rm LR}$),
local Andreev reflection on the left side (${\cal T}_{\rm A}$),
and cross Andreev reflection (${\cal T}_{\rm CA}$).
Full agreement is achieved between
the single-electron-wavefunction method (stationary limit)
and the nonequilibrium Green's function approach,
as shown by the corresponding curves and dots.   % }
(a)
Results based on simulation using the standard method of BdG equation
with both the positive and negative energy eigenstates
participating in the dynamics.
Parameters: $\mu=0$, $\Delta=0.5$, $t=1.0$,
and the tunnel coupling rates $\Gamma_L=\Gamma_R=\Gamma=0.05$.
(b)
Results from similar simulation as for (a),
except keeping only the positive energy eigenstates
by performing the projection as explained in the main text.    }
\end{figure}

\subsection{Results and Discussions}

Indeed, the SPWF approach has the particular advantage
to address time dependent transports. However, in this work
we restrict our interest to stationary results of the transport.

Before displaying our numerical results, we first quote the analytical results
based on the low-energy effective model
and the S-matrix scattering approach \cite{Flen19b,Bee08,Law09,Vu18,Gla02}.
Using the results derived in Ref.\ \cite{Bee08},
we obtain the local Andreev reflection, the cross Andreev reflection,
and the normal electron transmission coefficients
(${\cal T}_{\rm A}$, ${\cal T}_{\rm CA}$, and ${\cal T}_{\rm LR}$), respectively, as
\bea\label{T-eff}
{\cal T}_{\rm A}(E) &=& \Gamma^2_L(E^2+\Gamma^2_R) / |Z|^2 \,,  \nl
{\cal T}_{\rm CA}(E) &=& {\cal T}_{\rm LR}(E) = \epsilon^2_M \Gamma_L\Gamma_R / |Z|^2  \,,
\eea
where $Z=\epsilon^2_M-(E+i\Gamma_L)(E+i\Gamma_R)$.
The same results can be obtained as well using \Eq{nGF-2}, more straightforwardly.

In particular, under the limits of $\epsilon_M\to 0$ and $E\to 0$, we have
${\cal T}_{\rm A}\to 1$, being free from the coupling strength.
We notice that in Ref.\ \cite{Law09}, this type of full Andreev-reflection
(with unity coefficient) has been highlighted
in terms of Majorana-fermion-induced {\it resonant} Andreev reflection.
However, in Ref.\ \cite{Law09}, the local Andreev reflection is considered
for the setup where only one bound state of the Majorana pair
is coupled to the probe lead, while the other bound state is suspending
(without coupling to any probe lead).
This consideration corresponds to the setup of the standard two-probe
tunneling spectroscopy experiment, which probes the local Andreev reflection
taking place at the interface between a normal metal and {\it grounded} superconductor.
Actually, the resonant Andreev reflection with ${\cal T}_{\rm A}\to 1$
will result in the {\it quantized} zero-bias differential conductance,
$G=\frac{2e^2}{h}{\cal T}_{\rm A}\to 2 e^2/h$.
In this context, we may mention that for the local Andreev state,
or, the so-called quasi-Majorana states \cite{Vu18},
the one more Majorana state coupled to the same lead will result in
the conductance $G\to 4 e^2/h$, under certain parameter conditions.
The quantized conductance $2 e^2/h$ has been extensively analyzed
\cite{DS17,Vu18,DS01,Flen10,Flen16}
and was regarded as an important signature of Majorana states \cite{Kou18a}.

For the setup we consider here,
both sides of the Majorana wire are coupled to probing leads.
The fully `resonant' Andreev reflection on the left side
obtained also in this setup implies
that the electron-hole excitation at the left side
does not propagate to the other side,
since no coupling effect of the other side
is sensed in the probe of the local Andreev reflection.
Based on \Eq{T-eff}, we observe another remarkable feature, say,
under the limit $\epsilon_M\to 0$, ${\cal T}_{\rm CA}={\cal T}_{\rm LR}\to 0$.
This type of vanishing cross Andreev reflection and normal electron transmission
indicates also that the electron-hole excitations cannot propagate
from one side to the other through the Majorana quantum wire.

Indeed, all the above features (from the low-energy effective model)
are recovered in Fig.\ 4(a), by simulating
the electron and hole dynamics based on the Kitaev lattice model
% {\color{red}\bf
using both the SPWF and nGF approaches, by setting $\hat{P}=1$
which corresponds to the conventional BdG treatment.   % }
However, the results of the vanishing cross Andreev reflection
and normal electron transmission shown in Fig.\ 4(a)
are not consistent with the electron transfer dynamics
revealed from the simple `Dot-Wire-Dot' system
analyzed in Refs.\ \cite{Lee08,Fu10},
where the electron and hole excitations in the wire
(described by the occupied state $|n_f=1\ra$)
do correlate the two quantum dots and can
result in electron transfer and cross Andreev process between them.

In Fig.\ 4(b) we show the {\it consistent} results from new simulations,
based on the same Kitaev lattice model and
using both the SPWF and nGF approaches.
In the new simulations,
from the lesson learned earlier in the `Dot-Wire-Dot' setup, we allow
only coupling the electron and hole states of the leads
to the positive-energy Bogoliubov quasiparticles in the wire,
i.e., properly accounting for the projection of the wire states.
Remarkably, we find essential differences, compared to Fig.\ 4(a).
{\it (i)}
The transmission and cross Andreev reflection coefficients
are now {\it nonzero} in the limit $\epsilon_M\to 0$.
The basic reason is that in the projected Hilbert subspace
(after the action of the projector $\hat{P}$),
no `cancellation' of the electron-hole excitations occurs
on the right side of the quantum wire,
which yet would happen if including both the positive and negative
zero-energy eigengenstates in the naive treatment.
The results in Fig.\ 4(b) are now in agreement with
the teleportation picture revealed in Refs.\ \cite{Lee08,Fu10}.
$~~${\it (ii)}
For the local Andreev reflection (on the left side),
we find that the height of the reflection peak becomes 1/4,
rather than 1 as observed in Fig.\ 4(a).
We may understand this from the simplified low-energy effective
model of the single MZMs coupled to two probe leads.
Applying \Eq{nGF-2}, we have
\bea\label{nGF-3}
{\cal T}_A(E)=\Gamma^2_L / |E-\epsilon_M-i(\Gamma_L+\Gamma_R)|^2 \,.
\eea
Under the symmetric coupling to both leads ($\Gamma_L=\Gamma_R$),
we find ${\cal T}_A(E)\to 1/4$ when $E\to \epsilon_M$,
being also independent of the coupling strength.
However, if $\Gamma_L\neq \Gamma_R$, the result
is no longer independent of the coupling strengths.
We have examined this point as well by simulating the Kitaev lattice model.

As an extending discussion, let us consider to switch off the coupling
to the right lead, say, to set $\Gamma_R=0$.
We thus return to the situation considered in Ref.\ \cite{Law09}.
From \Eq{nGF-3}, as in Ref.\ \cite{Law09},
we also conclude that the resonant Andreev reflection coefficient is 1
and is independent of the coupling strength.
Again, this single-lead coupling corresponds to
the standard tunneling spectroscopy experiments
of detecting the Majorana zero modes
\cite{Kou12,Hei12,Fur12,Xu12,Fin13,Mar13,Mar16,Mar17,Kou18a},
and the coupling-strength-free resonant Andreev reflection
will result in the {\it Majorana quantized conductance} $2e^2/h$.
% {\color{red}\bf
However, the result will dramatically change if we consider a two-lead coupling device.
More specifically, following Ref.\ \cite{Bee08},
let us consider the two leads are equally voltage-biased
(with respect to the grounded superconductor),
and for simplicity assume a symmetric coupling to the two leads.
Then, based on the result of Fig.\ 4(b),
we obtain $G_{\rm A}=(\frac{2e^2}{h})(\frac{1}{4})=e^2/(2h)$,
by accounting for the contribution of the local Andreev reflection.
Moreover, for the equally biased two-lead setup,
the crossed Andreev reflection (which exists even at the limit $\epsilon_M\to 0$)
will contribute a conductance of
$G_{\rm CA}=(\frac{e^2}{h})(\frac{1}{4}+\frac{1}{4})=e^2/(2h)$.
Therefore, the total zero-bias-peak of the conductance probed at the left lead
is a sum of the two results above, i.e.,
$G=G_{\rm A}+G_{\rm CA}=e^2/h$, which is {\it a half}
of the popular value of the {\it Majorana conductance} ($2e^2/h$).
From the understanding based on Fig.\ 4(b) and \Eq{nGF-3}, we know that
this result manifests the nonlocal nature of the MZMs,
which allows both the crossed Andreev reflection (even at $\epsilon_M\to 0$)
and the `backward propagation' (to the left side)
of the self-energy effect owing to coupling to the right lead.   % }

% {\color{red}\bf
In the above analysis, we only considered the ideal case of $\epsilon_M\to 0$,
which is most dramatic for the issue of Majorana nonlocality.
If $\epsilon_M\neq 0$,
the insight gained from \Eq{nGF-3} indicates that the transmission peak
under the resonant condition $E\to \epsilon_M\neq 0$ is the same as $E\to \epsilon_M=0$.
This simply implies the same differential conductance
at the bias voltage $eV=\epsilon_M\neq 0$
as the zero-bias peak for $\epsilon_M=0$.
However, if we consider only the zero-bias case, which means $E\to 0$ (but $\epsilon_M\neq 0$),
we know from \Eq{nGF-3} that the transmission coefficient is lower than $1/4$
(for the symmetric coupling $\Gamma_L=\Gamma_R$),
which would result in a smaller zero-bias conductance.
Also, for the `Dot-Wire-Dot' setup considered in Sec.\ II,
if $\epsilon_M\neq 0$ but $\epsilon_1=\epsilon_2=\epsilon_M$,
the result of the resonant teleportation transfer is
the same as that from $\epsilon_1=\epsilon_2=\epsilon_M=0$.

However, as emphasized through the whole work,
if inserting the usual BdG treatment into the dynamics of
the charge transfer between the quantum dots
or transmission between two transport leads,
the vanishing energy $\epsilon_M = 0$ will vanish the charge transfer/transmission.
For $\epsilon_M \neq 0$, only the Rabi-transition type mechanism
will result in a state transfer between the MBSs
(a picture in the state basis of $|\gamma_1\ra$ and $|\gamma_2\ra$),
with the Rabi frequency given by the overlap energy $\epsilon_M$.
Nevertheless, this mechanism is fully different from
the transmission through the {\it single} Majorana energy level.        %  } %==

\section{Summary}

We have revisited the teleportation-channel-mediated
charge transfer and transport problems,
essentially rooted in the nonlocal nature of the MZMs.
We considered two setups:
the first one is a toy configuration,
say, a `quantum dot--Majorana wire--quantum dot' system,
while the second one is a more realistic transport setup
which is quite relevant to the tunneling spectroscopy experiments.
Through a simple analysis for the `teleportation' issue in the first setup,
we revealed a clear inconsistency between
the conventional BdG equation based treatment
and the method within the `second quantization' framework
(using the regular fermion number states of occupation).
We proposed a solving method to eliminate the discrepancy
and further considered the transport setup, by inserting the same spirit of treatment.
%%
% {\color{red}\bf
In this latter context, we developed a {\it single-particle-wavefunction}
approach to quantum transports, which renders both the conventional
quantum scattering theory and the steady-state nGF formalism
as its stationary limit.
We analyzed the tunneling conductance spectroscopy
for the Majorana two-lead coupling setup,
with comprehensive discussions and
a new prediction for possible demonstration by experiments.   %  } %===

\vspace{0.8cm}
%\begin{acknowledgements}
{\flushleft\it Acknowledgements.}---
This work was supported by the
National Key Research and Development Program of China
(No.\ 2017YFA0303304) and the NNSF of China (Nos.\ 11675016, 11974011 \& 11904261).
%\end{acknowledgements}

\end{CJK*}
\end{document}